\documentclass[10pt,twocolumn,twoside]{IEEEtran} 
\usepackage{amsmath}
\usepackage{array}
\usepackage{graphicx}

\begin{document}

\title{Deterministic Selection of Phase Sequences in Low Complexity SLM Scheme}

\author{Jun-Young Woo, Hyun-Seung Joo, Kee-Hoon Kim, Jong-Seon No, {\em Fellow, IEEE,} and Dong-Joon Shin, {\em Senior Member, IEEE}

\thanks{J.-Y. Woo, H.-S. Joo, and K.-H. Kim are with the Department of Electrical Engineering and Computer Science, INMC, Seoul
National University, Seoul,  Korea (e-mail: \{jywoo, joohs, kkh\}@ccl.snu.ac.kr).}
\thanks{J.-S. No is with the Department of Electrical Engineering and Computer Science, INMC, Seoul
National University, Seoul,~Korea (e-mail: jsno@snu.
ac.kr).}
\thanks{D.-J. Shin is with the Department of Electronic Engineering, Hanyang University, Seoul, Korea (e-mail: djshin@hanyang.ac.kr).}
}


\maketitle

\begin{abstract}
Selected mapping (SLM) is a suitable scheme, which can solve the peak-to-average power ratio (PAPR) problem. Recently, many researchers have concentrated on reducing the computational complexity of the SLM schemes. One of the low complexity SLM schemes is the Class III SLM scheme which uses only one inverse fast fourier transform (IFFT) operation for generating one orthogonal frequency division multiplexing (OFDM) signal sequence. By selecting rotations and cyclic shifts randomly, it can generate $N^3$ alternative OFDM signal sequences, where $N$ is the FFT size. But this selection can not guarantee the optimal PAPR reduction performances. Therefore, in this paper, we propose a simple deterministic cyclic shifts selection method which is optimal in case of having low variance of correlation coefficient between two alternative OFDM signal sequences. And we show that cyclic shifts are highly dependent on the PAPR reduction performance than rotations. For small FFT size and the number of alternative signal sequences is close to $N/8$, simulation results show that the proposed scheme can achieve better PAPR reduction performance than the Class III SLM scheme.
\end{abstract}


\setlength{\textfloatsep}{10pt}
\setlength{\dbltextfloatsep}{5pt}

\section{Introduction}\label{Introduction}
Orthogonal frequency division multiplexing (OFDM) is one of the most popular multicarrier modulation technique. Because of the orthogonality of subcarriers, the receiver can recover the transmitted data without interferences. Due to this robustness against the multipath channel, OFDM has been adopted as the standards for various wireless communication systems such as IEEE 802.11 a (WLAN), IEEE 802.16 (WiMAX), and long term evolution (LTE). But, it also has the serious drawback which is high peak-to-power ratio (PAPR). When OFDM signals pass through the high power amplifier (HPA), it has in-band distortion and out-of-band radiation. Thus, to reduce the PAPR, several schemes are proposed.

Clipping is the simplest technique, tone reservation  reserves the tones for peak canceling signal, and
probabilistic method, selected mapping (SLM) and partial transmit sequence (PTS), generate alternative signal sequences and choose the one with the lowest PAPR.

This paper is organized as follows. Section~\ref{sec:Class III SLM Scheme} introduces the Class III SLM scheme. Optimal condition and deterministic cyclic shifts generation method are proposed in Section~\ref{sec:Deterministic Selection of Phase Sequences}. In Section~\ref{sec:Simulation Result}, simulation results are shown and Section~\ref{sec:Conclusion} concludes this paper.

\section{Class III SLM Scheme}\label{sec:Class III SLM Scheme}
\begin{figure}[t]
  \center
  \includegraphics[width=9cm]{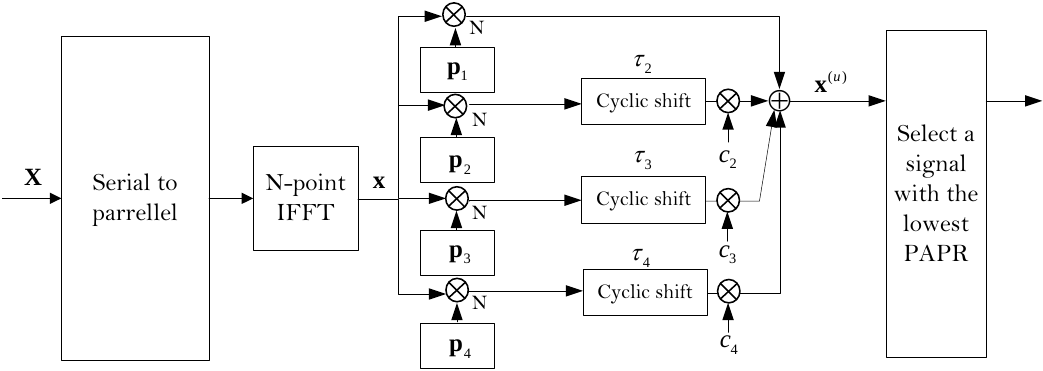}\\
  \caption{Block diagram of the Class III SLM scheme}\label{Fig:Visio-Class3_Wang}
\end{figure}

The Class III SLM scheme is introduced by Li \emph{et al.}~\cite{CLASS3} and it only uses one inverse fast fourier transform (IFFT)~operation for generating one orthogonal frequency division multiplexing (OFDM)~signal~sequence.~Fig.~\ref{Fig:Visio-Class3_Wang}~shows the block diagram of Class \mbox{III SLM} scheme.~Quadrature phase-shift keying (QPSK) or $M$-ary quadrature amplitude modulation ($M$-QAM) modulated input symbol sequence, ${\mathbf X}=[X_0,X_1,X_2,{\cdots},X_{N-1}]$, is the input of IFFT operation, where $N$ is the FFT size. Then, OFDM symbol sequence, ${\mathbf x}=[x_0,x_1,x_2,{\cdots},x_{N-1}]$, performs $N$-point circular convolution with $4$ base vectors.
The 4 base vectors are defined as
\begin{alignat}{2}\label{eq:4-base-vector}
\begin{split}
\mathbf{p}_1&=\left[\underbrace{1,0,{\cdots},0}_{\frac{N}{4}},\underbrace{1,0,{\cdots},0}_{\frac{N}{4}},\underbrace{1,0,{\cdots},0}_{\frac{N}{4}},\underbrace{1,0,{\cdots},0}_{\frac{N}{4}}\right]\\
\mathbf{p}_2&=\left[\underbrace{1,0,{\cdots},0}_\frac{N}{4},\underbrace{j,0,{\cdots},0}_{\frac{N}{4}},\underbrace{-1,0,{\cdots},0}_{\frac{N}{4}},\underbrace{-j,0,{\cdots},0}_{\frac{N}{4}}\right]\\
\mathbf{p}_3&=\left[\underbrace{1,0,{\cdots},0}_\frac{N}{4},\underbrace{-1,0,{\cdots},0}_\frac{N}{4},\underbrace{1,0,{\cdots},0}_\frac{N}{4},\underbrace{-1,0,{\cdots},0}_\frac{N}{4}\right]\\
\mathbf{p}_4&=\left[\underbrace{1,0,{\cdots},0}_\frac{N}{4},\underbrace{-j,0,{\cdots},0}_\frac{N}{4},\underbrace{-1,0,{\cdots},0}_\frac{N}{4},\underbrace{j,0,{\cdots},0}_\frac{N}{4}\right]\!\!.
\end{split}
\end{alignat}
Then, each sequence is right cyclic shifted, $\tau_i$, and rotated, $c_i$, to generate alternative signal sequences, ${\mathbf x}^{(u)}$. Finally, select the one with the lowest PAPR.
The conversion vector for generating $u$th alternative signal sequence has the general form
\begin{align}\label{eq:uth-conversion-vector}
    \mathbf{p}^{(u)}=\sum_{i=1}^{4}c_{i}^{(u)}{\mathbf{p}_{i}}_{\langle\tau_{i}^{(u)}\rangle}
\end{align}
where $0 \leq \tau_i\leq N/4-1$,~$c_i\in\{\pm1,\pm j\}$, $c_{i}^{(u)}$ is $i$th rotation value, and ${\mathbf{p}_{i}}_{\langle\tau_{i}^{(u)}\rangle}$ is $\tau_i^{(u)}$ right cyclic shifted version of $i$th base vector, $i=1,2,3$,~and~$4$.
By (\ref{eq:uth-conversion-vector}), the Class III SLM can generate $N^3$ alternative signal sequences.
\begin{figure}[t]
  \center
  \includegraphics[width=9cm]{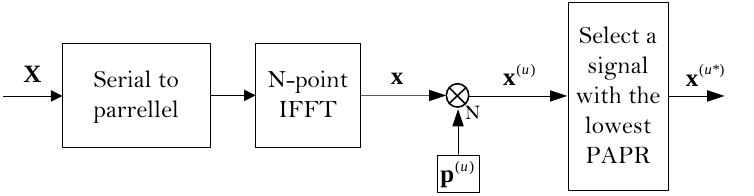}\\
  \caption{Block diagram of equivalent Class III SLM scheme}\label{Fig:Visio-Class3_Equivalent}
\end{figure}
Fig.~\ref{Fig:Visio-Class3_Equivalent} shows the equivalent Class III SLM scheme which is expressed using (\ref{eq:uth-conversion-vector}) and it is identical to Fig.~\ref{Fig:Visio-Class3_Wang}.

\section{Deterministic Selection of Cyclic Shifts}\label{sec:Deterministic Selection of Phase Sequences}

\subsection{Correlation Analysis}\label{sec:Correlation Analysis}

Heo {\em et al.}~\cite{Heo} showed that the correlation coefficient, $\rho_{ij}$, between sample powers of two alternative OFDM signal sequences is approximated as follows
\begin{align}\label{eq:rho}
    \rho_{ij}(\tau)\!\simeq\!\frac{1}{N^2}\left|\sum_{k=0}^{N-1}P^{(i)}(k)P^{(j)*}(k)e^{-j\frac{2\pi k}{N}\tau}\right|^2
\end{align}
where $P^{(i)}(k)$ is $k$th element in $i$th phase sequence, $(\cdot)^*$ denotes complex conjugate operation, and $0\leq\tau\leq N-1$. Let ${\mathbf P}_1=\rm FFT\{{\mathbf p}_1\}$. Then, FFT outputs of 4 base vectors in (\ref{eq:4-base-vector}) are expressed as
\begin{align}
\label{eq:4-subphase-vector}
\begin{split}
\mathbf{P}_1&=4\left[\underbrace{1,0,0,0}_{4},\underbrace{1,0,0,0}_{4},{\cdots},\underbrace{1,0,0,0}_{4}\right]\\
\mathbf{P}_2&=4\left[\underbrace{0,1,0,0}_{4},\underbrace{0,1,0,0}_{4},{\cdots},\underbrace{0,1,0,0}_{4}\right]\\
\mathbf{P}_3&=4\left[\underbrace{0,0,1,0}_{4},\underbrace{0,0,1,0}_{4},{\cdots},\underbrace{0,0,1,0}_{4}\right]\\
\mathbf{P}_4&=4\left[\underbrace{0,0,0,1}_{4},\underbrace{0,0,0,1}_{4},{\cdots},\underbrace{0,0,0,1}_{4}\right]\!\!.
\end{split}
\end{align}
Because circular convolution in time domain is identical to component wise multiplication in frequency domain, $\mathbf{x}\oplus_N\mathbf{p}_1=\mathbf{X}\odot\mathbf{P}_1$, input symbol sequence is multiplied by $\mathbf{P}_i, i=1,2,3,$ and $4$. It means that the 4 vectors, $\mathbf{P}_i$, are partitioning vectors same as the partitioning in the PTS. For example, $\mathbf{P}_1$ has all zero elements except when its indices are $0$ mod $4$. Thus, $\mathbf{X}\odot\mathbf{P}_1=[X_0,0,0,0,X_4,0,0,0,{\cdots},X_{N-4},0,0,0]$. It can also be applied to other partitioning vectors. After all, the Class III SLM scheme can be interpreted as an interleaved partitioned PTS scheme \cite{Interleaved-Partition}.

$u$th alternative signal sequence in the Class III SLM scheme can be expressed as
\begin{align}\label{eq:uth-alternative-sequence}
\mathbf{x}^{(u)}=&~\mathbf{p}^{(u)}\otimes_N\mathbf{x}\nonumber\\
                =&\left(\sum_{i=1}^{4}c_{i}^{(u)}{\mathbf{p}_{i}}_{\langle\tau_{i}^{(u)}\rangle}\right)\otimes_N\mathbf{x}\nonumber\\
                =&~{\rm IFFT}\left\{\left(\sum_{i=1}^{4}c_{i}^{(u)}{\mathbf{P}_{i}}^{\left(\tau_{i}^{(u)}\right)}\right)\odot\mathbf{X}\right\}
\end{align}
where ${\mathbf{P}_{i}}^{(\tau_{i}^{(u)})}$ denotes the linear phase in frequency domain by cyclic shift, $\tau_{i}^{(u)}$, in time domain.~And then, we can consider $\sum_{i=1}^{4}c_{i}^{(u)}{\mathbf{P}_{i}}^{(\tau_{i}^{(u)})}$ in (\ref{eq:uth-alternative-sequence}) as a phase sequence in SLM scheme. Thus, $u$th phase sequence in the Class III SLM scheme can be expressed as follows
\begin{align}\label{eq:uth-phase-sequence}
\mathbf{P}^{(u)}=\sum_{i=1}^{4}c_{i}^{(u)}{\mathbf{P}_{i}}^{\left(\tau_{i}^{(u)}\right)}.
\end{align}
For $i=1$ in (\ref{eq:uth-phase-sequence}), it can be computed as
\begin{align}\label{eq:1mod4-element-phase-sequence}
c_{1}^{(u)}{\mathbf{P}_{1}}^{\left(\tau_{1}^{(u)}\right)}
\!\!=\!\!\left[\underbrace{c_{1}^{(u)},0,0,0}_{4},
{\cdots},
\underbrace{c_{1}^{(u)}e^{-j\frac{2\pi (N-4)\tau_{1}^{(u)}}{N}},0,0,0}_{4}\right]\!\!.
\end{align}
\begin{figure*}[ht]
\begin{align}\label{eq:uth-phase-sequence-all-element}
\mathbf{P}^{(u)}\!\!&=\!\sum_{i=1}^{4}c_{i}^{(u)}{\mathbf{P}_{i}}^{\left(\tau_{i}^{(u)}\right)}\nonumber\\
 &=\!\!\Bigg[c_{1}^{(u)}\!\!,c_{2}^{(u)}\!e^{-j\frac{2\pi \tau_{2}^{(u)}\!\!\!\!}{N}}\!\!,c_{3}^{(u)}\!e^{-j\frac{2\pi 2\tau_{3}^{(u)}\!\!\!\!}{N}}\!\!,c_{4}^{(u)}\!e^{-j\frac{2\pi 3\tau_{4}^{(u)}\!\!\!\!}{N}}\!\!\!,{\cdots},c_{1}^{(u)}\!e^{-j\frac{2\pi (N-4)\tau_{2}^{(u)}\!\!\!\!}{N}}\!\!,c_{2}^{(u)}\!e^{-j\frac{2\pi (N-3)\tau_{2}^{(u)}\!\!\!\!}{N}}\!\!,c_{3}^{(u)}\!e^{-j\frac{2\pi (N-2)\tau_{3}^{(u)}\!\!\!\!}{N}}\!\!,c_{4}^{(u)}\!e^{-j\frac{2\pi (N-1)\tau_{4}^{(u)}\!\!}{N}}\Bigg]\nonumber\\ \nonumber\\ \hline
\end{align}
\end{figure*}
With this result, we can compute the $u$th phase sequence as (\ref{eq:uth-phase-sequence-all-element}), where the elements are
\begin{equation}\label{eq:uth-phase-sequence-element}
P^{(u)}(k)=
\left\{
  \begin{array}{ll}
    c_{1}^{(u)}e^{-j\frac{2\pi k\tau_{1}^{(u)}}{N}}, &\quad k=0~{\rm mod}~4 \\
    c_{2}^{(u)}e^{-j\frac{2\pi k\tau_{2}^{(u)}}{N}}, &\quad k=1~{\rm mod}~4 \\
    c_{3}^{(u)}e^{-j\frac{2\pi k\tau_{3}^{(u)}}{N}}, &\quad k=2~{\rm mod}~4\\
    c_{4}^{(u)}e^{-j\frac{2\pi k\tau_{4}^{(u)}}{N}}, &\quad k=3~{\rm mod}~4.
  \end{array}
\right.
\end{equation}
Finally, we can substitute (\ref{eq:uth-phase-sequence-all-element}) to (\ref{eq:rho}) and then consider only inner parts of the magnitude $|\cdot|$. We can replace it as ${\rm A}(\tau)$ and then we can get 4 terms. We can also replace each term with ${\rm A}_{(i)}(\tau)$, $i=1$, 2, 3, and 4, as follows
\begin{align}\label{eq:A(tau)}
{\rm A}(\tau)=&~\sum_{k=0}^{N-1}\textrm{P}^{(i)}(k)\textrm{P}^{(j)*}(k)e^{-j\frac{2\pi k}{N}\tau}\nonumber\\
=&~\sum_{v=0}^{\frac{N}{4}-1} c_{1}^{(i)}c_{1}^{(j)*}e^{-j\frac{2\pi(4v) }{N}\left(\tau+\tau_{1}^{(i)}-\tau_{1}^{(j)}\right)}\nonumber\\
+&~\sum_{v=0}^{\frac{N}{4}-1} c_{2}^{(i)}c_{2}^{(j)*}e^{-j\frac{2\pi(4v+1) }{N}\left(\tau+\tau_{2}^{(i)}-\tau_{2}^{(j)}\right)}\nonumber\\
+&~\sum_{v=0}^{\frac{N}{4}-1} c_{3}^{(i)}c_{3}^{(j)*}e^{-j\frac{2\pi(4v+2) }{N}\left(\tau+\tau_{3}^{(i)}-\tau_{3}^{(j)}\right)}\nonumber\\
+&~\sum_{v=0}^{\frac{N}{4}-1} c_{4}^{(i)}c_{4}^{(j)*}e^{-j\frac{2\pi(4v+3) }{N}\left(\tau+\tau_{4}^{(i)}-\tau_{4}^{(j)}\right)}\nonumber\\
=&~{\rm A}_{(1)}(\tau)+{\rm A}_{(2)}(\tau)+{\rm A}_{(3)}(\tau)+{\rm A}_{(4)}(\tau).
\end{align}
Let us consider one ${\rm A}_{(i)}(\tau)$. And take the rotation $c_{1}^{(i)}c_{1}^{(j)*}$ out of summation, then we can get ${\rm \bar{A}}_{(1)}(\tau)$.
\begin{align}\label{eq:A(1)(tau)}
{\rm A}_{(1)}(\tau)=&~\sum_{v=0}^{\frac{N}{4}-1} c_{1}^{(i)}c_{1}^{(j)*}e^{-j\frac{2\pi(4v) }{N}\left(\tau+\tau_{1}^{(i)}-\tau_{1}^{(j)}\right)}\nonumber\\
=&~c_{1}^{(i)}c_{1}^{(j)*}\sum_{v=0}^{\frac{N}{4}-1}e^{-j\frac{2\pi(4v) }{N}\left(\tau+\tau_{1}^{(i)}-\tau_{1}^{(j)}\right)}\nonumber\\
=&~c_{1}^{(i)}c_{1}^{(j)*}\sum_{v=0}^{\frac{N}{4}-1}e^{-j\frac{2\pi(v) }{N/4}\left(\tau+\tau_{1}^{(i)}-\tau_{1}^{(j)}\right)}\nonumber\\
=&~c_{1}^{(i)}c_{1}^{(j)*}{\rm \bar{A}}_{(1)}(\tau).
\end{align}
In the same manner, the rest can also be computed as
\begin{align}\label{eq:A(2,3,4)-bar-(tau)}
   {\rm \bar{A}}_{(2)}(\tau)=&\sum_{v=0}^{\frac{N}{4}-1}e^{-j\frac{2\pi(v+1/4) }{N/4}\left(\tau+\tau_{2}^{(i)}-\tau_{2}^{(j)}\right)}\nonumber\\
   {\rm \bar{A}}_{(3)}(\tau)=&\sum_{v=0}^{\frac{N}{4}-1}e^{-j\frac{2\pi(v+2/4) }{N/4}\left(\tau+\tau_{3}^{(i)}-\tau_{3}^{(j)}\right)}\\
   {\rm \bar{A}}_{(4)}(\tau)=&\sum_{v=0}^{\frac{N}{4}-1}e^{-j\frac{2\pi(v+3/4) }{N/4}\left(\tau+\tau_{4}^{(i)}-\tau_{4}^{(j)}\right)}\nonumber.
\end{align}
If the value of $\big(\tau+\tau_{1}^{(i)}-\tau_{1}^{(j)}\big)$ is multiple of $N/4$, ${\rm \bar{A}}_{(1)}(\tau)$ becomes $N/4$. Otherwise, it becomes zero since the sum of all points in an unit circle is zero. Let $d_{\tau_v}=\tau_v^{(i)}-\tau_v^{(j)}$, $v=1,2,3$, and $4$, be the difference between two cyclic shifts. Then, we can calculate all possible values of ${\rm \bar{A}}_{(i)}(\tau)$ with different $\tau$. The results are shown in Table~\ref{table:A-tau}. ${\rm \bar{A}}_{(i)}(\tau)$ can have 4 possible values each. Therefore, if four ${\rm \bar{A}}_{(i)}(\tau)$ have values without sharing $\tau$, there can be 16 values for different $\tau$.

\renewcommand{\arraystretch}{2}
\begin{table}[t]
\centering\caption{${\rm \bar{A}}_{(i)}(\tau)$ values for different $\tau$.}
\begin{tabular}{|c!{\vrule width 1.5pt}c|c|c|c|}
\hline
$\tau$ & $-d_{\tau_i}~{\rm mod}~N$ & $\frac{N}{4}-d_{\tau_i}$ & $\frac{2N}{4}-d_{\tau_i}$ & $\frac{3N}{4}-d_{\tau_i}$ \\ \hline\hline
${\rm \bar{A}}_{(1)}(\tau)$ & $\frac{N}{4}$ & $\frac{N}{4}$ & $\frac{N}{4}$ & $\frac{N}{4}$ \\ \hline
${\rm \bar{A}}_{(2)}(\tau)$ & $\frac{N}{4}$ & $\frac{N}{4}j$ & $\frac{N}{4}$ & $-\frac{N}{4}j$ \\ \hline
${\rm \bar{A}}_{(3)}(\tau)$ & $\frac{N}{4}$ & $-\frac{N}{4}$ & $\frac{N}{4}$ & $-\frac{N}{4}$ \\ \hline
${\rm \bar{A}}_{(4)}(\tau)$ & $\frac{N}{4}$ & $-\frac{N}{4}j$ & $\frac{N}{4}$ & $\frac{N}{4}j$ \\
\hline
\end{tabular}
\label{table:A-tau}
\end{table}
\subsection{Optimal condition for low variance of correlation}\label{sec:Condition for Good Phase Sequences}
In~\cite{VarianceofCorrelation}, authors showed that phase sequence sets with low variance of correlation have better PAPR reduction performance. Based on this results, we found the condition which can make the variance of correlation coefficient low.
\begin{equation}\label{eq:optimal-condition}
    {\rm Optimal~condition}:~d_{\tau_1}\neq d_{\tau_2}\neq d_{\tau_3}\neq d_{\tau_4}~{\rm mod}~\frac{N}{4}.
\end{equation}
That is, to achieve optimal PAPR reduction performance, we have to choose the cyclic shifts with different $d_{\tau_v}$ between any ${U \choose 2}$ alternative signal sequence pairs, where $U$ is the total number of alternative signal sequences. Suppose 2 or more ${\rm \bar{A}}_{(i)}(\tau)$ have the values in Table~\ref{table:A-tau} at same $\tau'$. By Parseval's Theorem, the magnitude should be larger than a case which have only one signal at $\tau'$. Therefore, sharing same $\tau'$ might increase the variance of correlation and it is not desirable \cite{VarianceofCorrelation}.

\subsection{Deterministic Cyclis Shifts Generation Method}\label{sec:Deterministic Generation Method}
In this subsection, we propose a simple cyclic shifts generation method. Table \ref{table:Deterministic-generation-of-phase-sequences} shows the cyclic shifts for each $\tau_i$, where $u$ denotes the alternative signal sequence index. $\tau_2$ has the values which are multiple of 1 at an increasing order. Similarly, $\tau_3$ has multiple of 2 and $\tau_4$ has multiple of 3. For any $u_1,u_2$, where $u_1\neq u_2$ and $u_1<u_2$, the corresponding differences are  $d_{\tau_1}=0$, $d_{\tau_2}=u_2-u_1$, $d_{\tau_3}=2(u_2-u_1)$, and $d_{\tau_4}=3(u_2-u_1)=2(u_2-u_1)+(u_2-u_1)$, respectively. If $u_2-u_1=\frac{N}{8}$, $d_{\tau_2}=d_{\tau_4}$ mod $\frac{N}{4}$. Therefore, $u_2-u_1<\frac{N}{8}$. Now, we have show that $d_{\tau_i}-d_{\tau_j}=k(u_2-u_1)\neq0$ mod $\frac{N}{4}$, where $i,j\in\{1,2,3,4\}, i<j$, and $k=1,2,$ and $3$. When $k=1$ and $2$, this condition holds by $u_1<u_2$ and $u_2-u_1<\frac{N}{8}$. And when $k=3$, because of $N=2^m$, where $m$ is natural number, $3(u_2-u_1)\neq0~{\rm mod}~\frac{N}{4}=2^{m-2}$. Thus, when $u_2-u_1=\frac{N}{8}$, it always satisfy the optimal condition and the maximum number of alternative signal sequences is $\frac{N}{8}$.
\begin{table}[t]
\centering\caption{Deterministic generation of phase sequences.}
\begin{tabular}{|c!{\vrule width 1.5pt}c|c|c|c|c|}
  \hline
  $u_i$ & $1$ & $2$ & $3$ & ${\cdots}$ & $k$  \\ \hline\hline
  $\tau_2$ & $1$ & $2$ & $3$ & ${\cdots}$ & $k$~mod~$\frac{N}{4}$ \\ \hline
   $\tau_3$ & $2$ & $4$ & $6$ & ${\cdots}$ & $2k$~mod~$\frac{N}{4}$ \\ \hline
  $\tau_4$ & $3$ & $6$ & $9$ & ${\cdots}$ & $3k$~mod~$\frac{N}{4}$ \\ \hline
\end{tabular}
\label{table:Deterministic-generation-of-phase-sequences}
\end{table}

\begin{figure}[t]
\center
  \includegraphics[width=8cm]{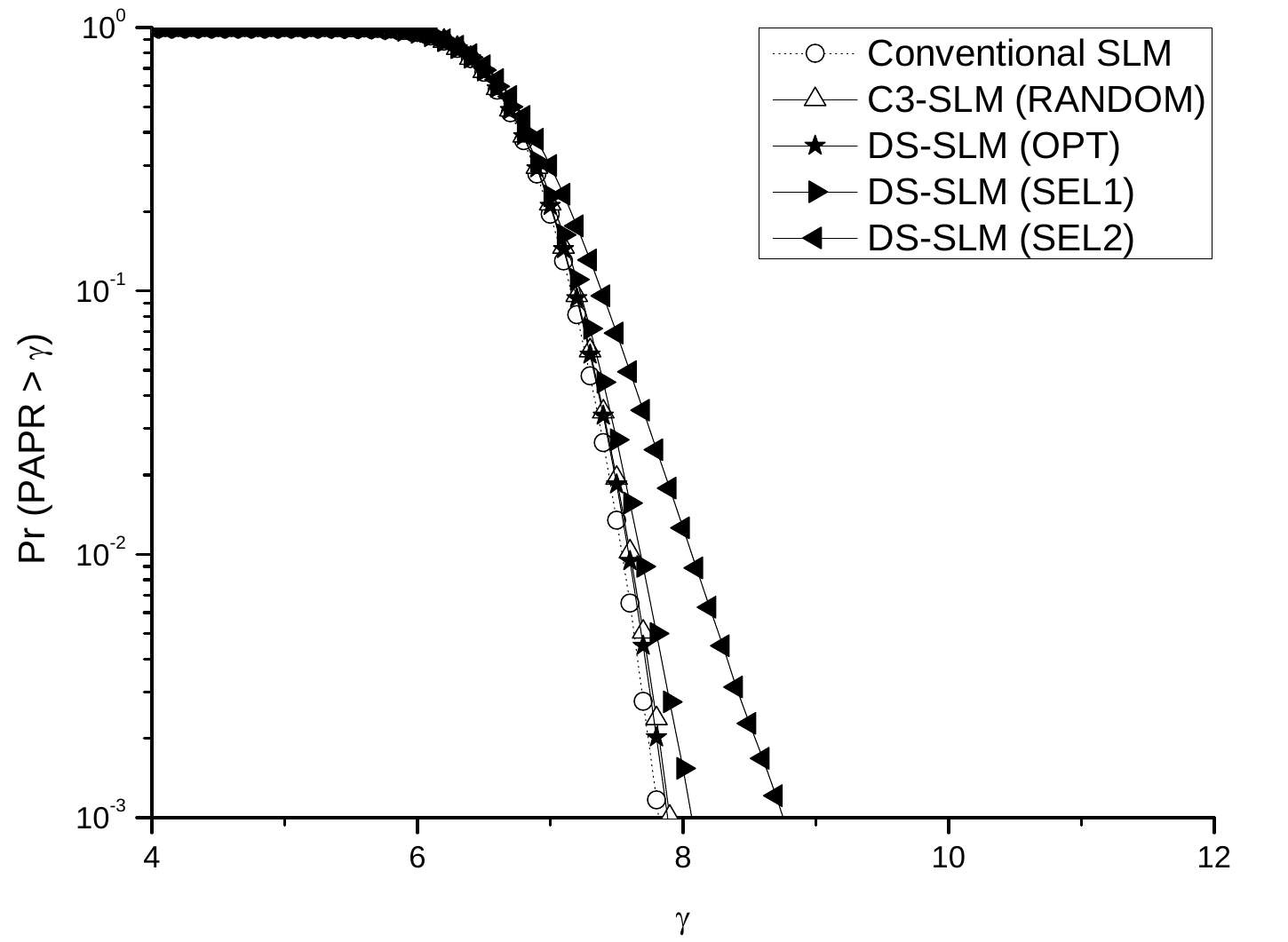}\\
  \caption{PAPR reduction performance when 16-QAM, $N=256$, and $U=10$.}\label{fig:PAPR-N256-U10}
\end{figure}
\begin{figure}[t]
\center
  \includegraphics[width=8cm]{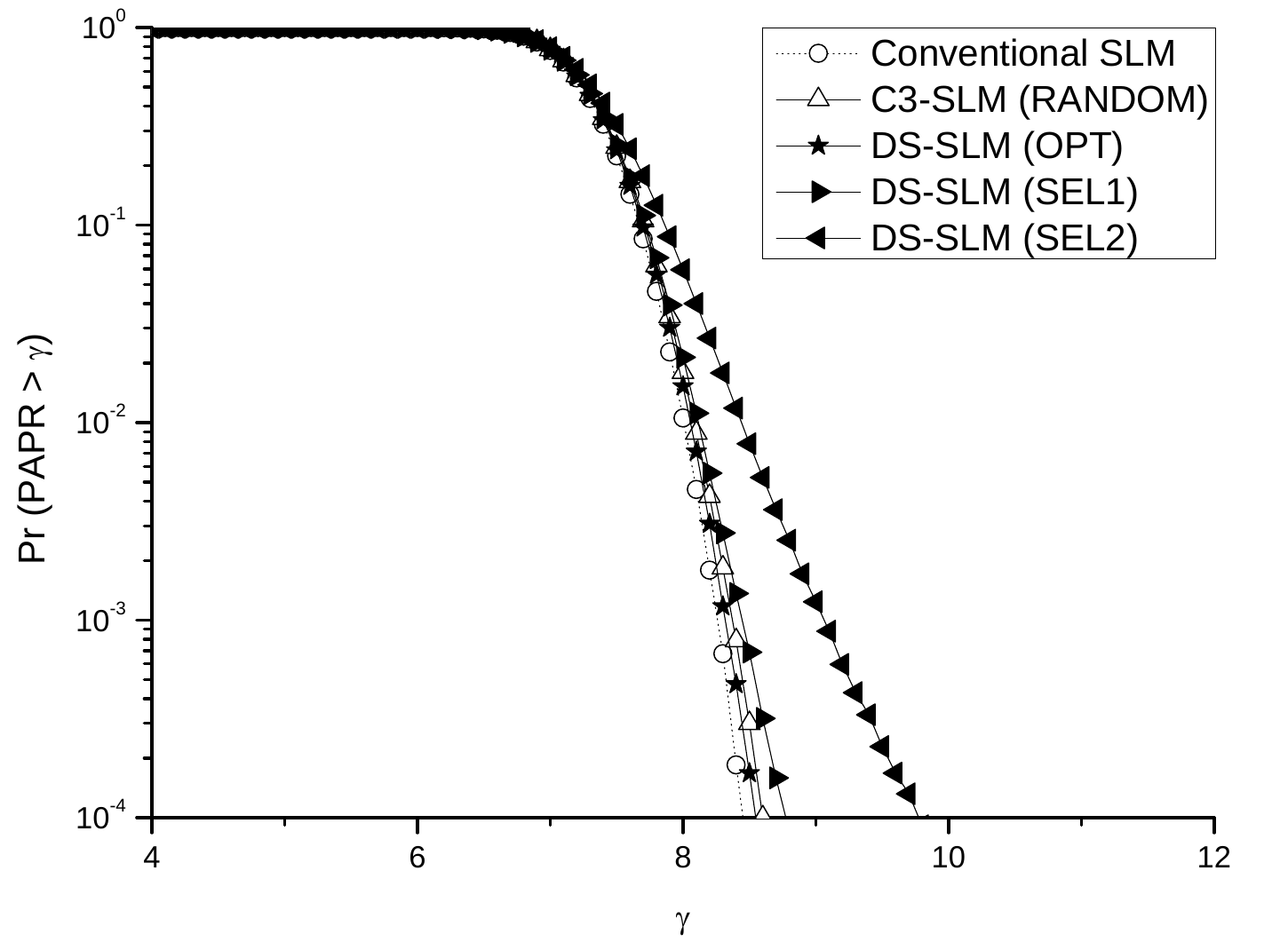}\\
  \caption{PAPR reduction performance when 16-QAM, $N=512$, and $U=10$.}\label{fig:PAPR-N512-U10}
\end{figure}
\begin{figure}[t]
\center
  \includegraphics[width=8cm]{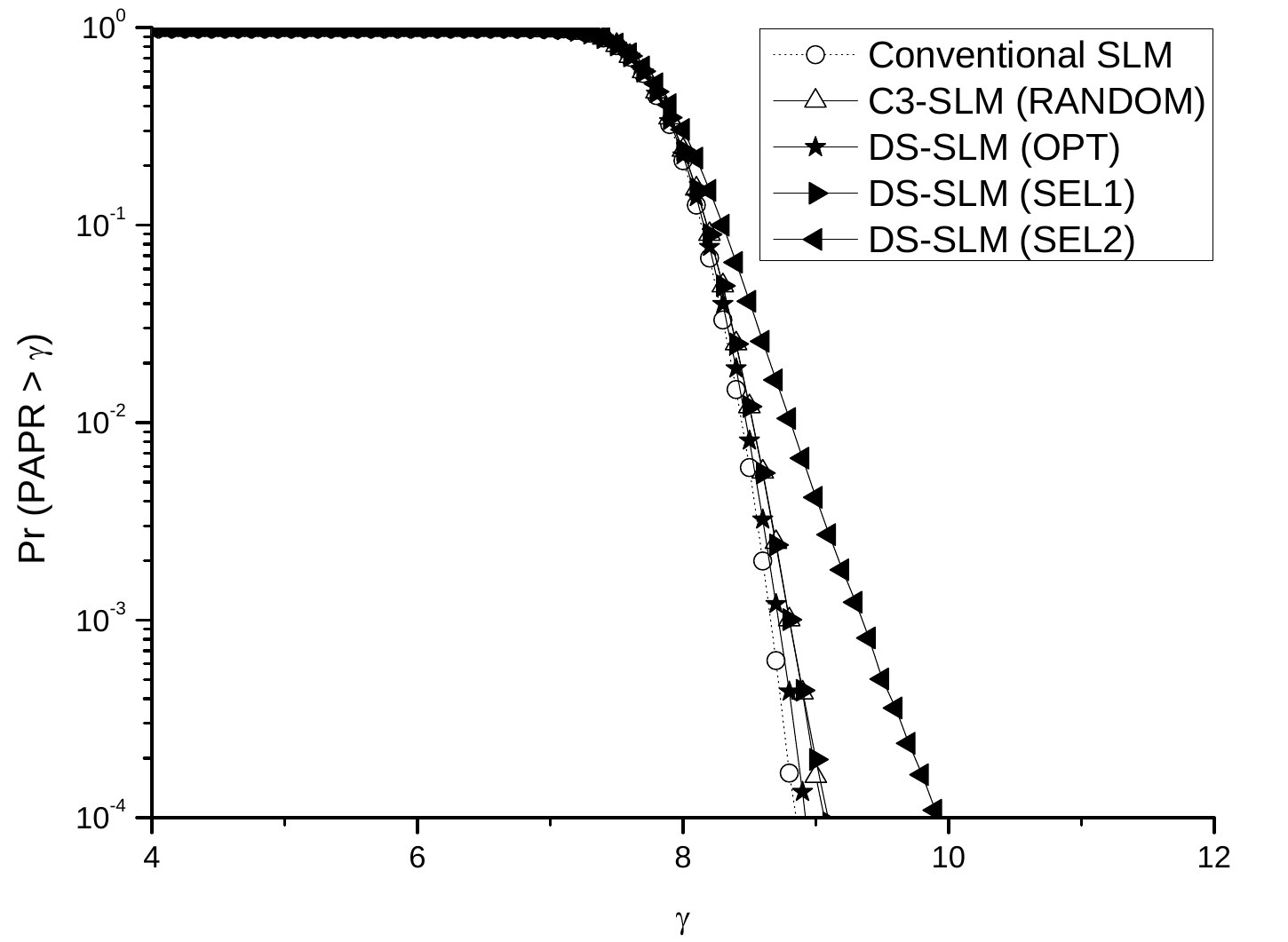}\\
  \caption{PAPR reduction performance when 16-QAM, $N=1024$, and $U=10$.}\label{fig:PAPR-N1024-U10}
\end{figure}

\section{Simulation Result}\label{sec:Simulation Result}
In this section, we compare the PAPR reduction performance between Class III SLM (C3-SLM) scheme and the proposed deterministic selection SLM scheme (DS-SLM). PAPR reduction performance is evaluated by complementary cumulative distribution function (CCDF) which shows a probability that is larger than a certain threshold level, $\gamma$. In C3-SLM scheme, cyclic shift and rotation values are randomly generated. While in DS-SLM scheme, cyclic shift values are taken from Table~\ref{table:Deterministic-generation-of-phase-sequences} and rotation values are all 1, that is, $c_2=c_3=c_4=1$.~Also, \mbox{$d_{\tau_1}= d_{\tau_2}$ and $d_{\tau_1}= d_{\tau_2}= d_{\tau_3}$} cases are considered for comparison. We call the C3-SLM with random generation of rotation and cyclic shift values as C3-SLM (RANDOM), DS-SLM with optimal condition as DS-SLM (OPT), DS-SLM with $d_{\tau_1}= d_{\tau_2}$ case as DS-SLM (SEL1), and  DS-SLM with $d_{\tau_1}= d_{\tau_2}= d_{\tau_3}$ case as DS-SLM (SEL2), respectively.
Figs.~(\ref{fig:PAPR-N256-U10})--(\ref{fig:PAPR-N1024-U10}) show the PAPR reduction performance of C3-SLM and DS-SLM scheme, when 16-QAM, $U=10$, and with different FFT size. Conventional SLM has the best PAPR reduction performance in all figures. Because of the correlation in C3-SLM scheme, it can not achieve better PAPR reduction performance than conventional SLM. When $N=256$, C3-SLM (RANDOM) and DS-SLM (OPT) can achieve almost same PAPR reduction performance. Since DS-SLM (SEL1) and DS-SLM (SEL2) can not satisfy the optimality condition, PAPR reduction performances are degraded than DS-SLM (OPT). This result shows that $d_{\tau_i}$ is highly dependent on the PAPR reduction performance than rotation values. When $N=512$, DS-SLM (OPT) can achieve better PAPR reduction performance than C3-SLM (RANDOM) slightly, and when $N=1024$, C3-SLM (RANDOM) has almost same PAPR reduction performance as DS-SLM (SEL1) which is degraded than DS-SLM (OPT). Therfore, when $N$ is large, DS-SLM (OPT) can achieve better PAPR reduction performance than C3-SLM (RANDOM).

\section{Conclusion}\label{sec:Conclusion}
 In this paper, we propose a simple deterministic cyclic shifts generation method which satisfy the optimal condition and show that cyclic shifts are highly dependent on the PAPR reduction performance than rotations. Simulation results show that the proposed DS-SLM scheme can achieve better PAPR reduction performance than the C3-SLM scheme for small FFT size and $U$ is close to $N/8$.

\section*{Acknowledgment}
This work was supported by the National Research Foundation of Korea (NRF) grant funded by the Korea government (MEST) (No. 2012-0000186).

\end{document}